\documentclass{IEEEarxiv}

\usepackage[colorlinks,urlcolor=blue,linkcolor=blue,citecolor=blue]{hyperref}
\expandafter\def\expandafter\UrlBreaks\expandafter{\UrlBreaks\do\/\do\*\do\-\do\~\do\'\do\"\do\-}
\usepackage{upmath,color}

\usepackage{doi}

\PassOptionsToPackage{hyphens}{url}

\def\comment#1{\mbox{}}
\jvol{\comment{XX}}
\jnum{\comment{XX}}
\paper{\comment{8}}
\jmonth{March}
\jname{\textit{\comment{}}}
\jtitle{\textit{\comment{}}}
\pubyear{2026}

\begin{document}

\sptitle{People of Computing}

\title{Jean-Raymond Abrial: 
A Scientific Biography of~a Formal~Methods~Pioneer}

\author{Jonathan P.\ Bowen}
\affil{\mbox{}London South Bank University, London SE1 0AA, UK}

\author{Henri Habrias}
\affil{University of Nantes, Nantes 44312, France}

\markboth{PEOPLE OF COMPUTING}{PEOPLE OF COMPUTING}

\begin{abstract}\looseness-1%
Jean-Raymond Abrial is one of the central figures in the
development of formal methods for software and systems
engineering. Over a career spanning more than five decades, he
has played a decisive role in the creation of the Z
specification notation, the B-Method, and Event-B, and in
demonstrating their applicability to large-scale industrial
systems. This paper presents a scholarly biographical account of
Abrial's life and work, tracing the evolution of his ideas from
early work on real-time languages and databases, through
foundational contributions to formal specification, refinement,
and proof, to the development of industrial-strength tool
support such as the Atelier~B and the Rodin platform. The paper
situates Abrial's contributions within their historical,
intellectual, and industrial contexts, and assesses their
lasting impact on software engineering and formal reasoning
about programs.
\end{abstract}

\maketitle

\chapteri{T}he history of formal methods in software engineering
is inseparable from the work of Jean-Raymond Abrial. While many
researchers have contributed to the theoretical foundations or
to isolated industrial applications, Abrial stands out for
having consistently pursued a single, coherent scientific
objective: to make the construction of correct software systems,
a disciplined, mathematically founded engineering activity. As
Abrial himself has remarked, he belongs to the class of
``monomaniac'' researchers who devote their careers to one
fundamental problem pursued from multiple angles over
time~\cite{JRA2015a}.

This article provides a biographical narrative of Abrial's career
and scientific contributions, drawing in particular on the
extensive historical and documentary material collected by Henri
Habrias \cite{Habrias2026}, as well as on Abrial's own
publications and retrospective accounts (see Appendix). The
emphasis is not only on chronology, but also on the evolution of
ideas: from early experience with real-time systems and
programming languages, through the invention of Z, to the
maturation of refinement and proof in the B and Event-B methods.

\begin{figure}
\centerline{\includegraphics[width=18.5pc]{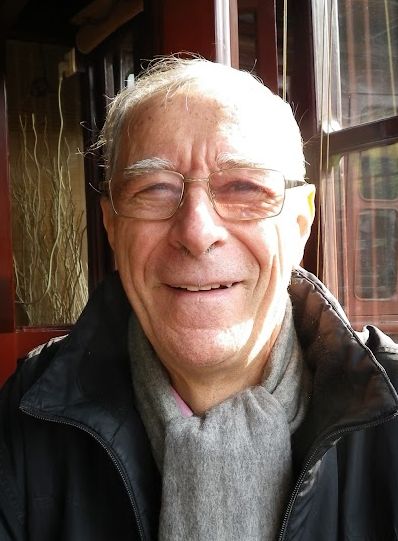}}
\caption{Photograph of Jean-Raymond Abrial in Shanghai, China, 2012.
(Cropped from Wikipedia \cite{Wikipedia}, used with permission.)}
\vspace*{-5pt}
\end{figure}

\section{EARLY LIFE AND EDUCATION}

Jean-Raymond Abrial was born on 6~November~1938 in Versailles,
France \cite{Habrias2026}. He received part of his early
education at the Prytanée National Militaire de La Flèche, a
prestigious French military school. In 1958, he entered the
École Polytechnique, from which he graduated in 1960 as an
engineer of the Génie Maritime.

Unlike many academic contemporaries, Abrial did not pursue a
doctoral thesis. This choice, which he later discussed
candidly, reflected both personal inclination and the
professional context of the time: his early career was oriented
toward applied research and large-scale engineering projects
rather than academic credentialing. This pragmatic orientation
would remain a defining feature of his approach to formal
methods.

\section{EARLY CAREER: REAL-TIME SYSTEMS AND PROGRAMMING LANGUAGES}

After graduating from the École Polytechnique, Abrial joined the
Centre de Programmation de la Marine (CPM), where he worked on
real-time systems for the French Navy. One of his most
significant early contributions was to the development of the
real-time programming language LTR-2 (Langage Temps R\'eel),
used in naval combat systems such as SENIT~\cite{JRA1966}.

This period exposed Abrial to the realities of safety-critical
and mission-critical software long before such terminology
became common. He encountered firsthand the difficulty of
reasoning about the correctness of complex concurrent systems
written in low-level or ad~hoc languages. These experiences
planted the seeds for his later insistence on rigorous
specification and proof.

Between 1963 and 1965, Abrial studied in the United States at
Stanford University as a French government scholar. There he
acquired a broad and systematic view of computer science, which
complemented his practical engineering background and exposed
him to emerging theoretical perspectives. At this time, John
McCarthy was in the Computer Science Division at Stanford, so it
is likely that Abrial would have been exposed to the
mathematically oriented LISP programming language. ALGOL was
also taught as an Advanced Computer Programming course for
graduate students and the programming language pioneer Niklaus
Wirth joined as an assistant professor of computer science at
Stanford too. This environment is likely to have had a
significant influence on Abrial's interests in programming and
specification languages, together with their semantics.

\section{DATABASES AND DATA SEMANTICS}

In the late 1960s and early 1970s, Abrial played a key role in
the development of the \emph{Socrate} database system
\cite{JRA1970a}, later commercialized as CLIO. Although he did
not initially specialize in database systems, Abrial introduced
several innovative ideas, including virtual addressing,
network-based data organization, and user-oriented data access
languages~\cite{Habrias2026}.

His theoretical reflections on these systems culminated in the
seminal paper \emph{Data Semantics} \cite{JRA1974},
which offered a rigourous, mathematical treatment of the meaning
of data structures. This work is widely regarded as a
foundational contribution to the formal semantics of databases
and foreshadowed Abrial's later work on formal specification.

\section{THE Z SPECIFICATION LANGUAGE}

Abrial's involvement in the design of the Ada programming
language in the 1970s marked a turning point \cite{JRA1983}. As
a member of the ``Green'' team led by Jean Ichbiah, Abrial
became deeply concerned with the gap between informal
requirements and formal programs. Natural-language explanations,
he observed, could not be reliably compared with executable
code.

During a research position at Oxford University (1979--1981),
invited by the leader of the Programming Research Group there,
Tony Hoare, Abrial developed a formal specification notation
based on typed set theory and first-order logic with Bernard
Sufrin and others \cite{Sufrin2024}. This formal
notation became known as Z \cite{Spivey1989}. The Z notation was
not conceived as an executable programming language, but as a
precise, mathematically grounded language for describing system
properties abstractly and unambiguously.

Abrial's teaching and internal reports at Oxford, together with
collaborations involving researchers such as Sufrin, Steve
Schuman and Bertrand Meyer \cite{Sufrin2024}, laid the
foundations for Z, as it would later be disseminated and
popularized \cite{JRA1980c}.  Z rapidly became one of the most
influential formal specification languages of the 1980s and
1990s, being taught on many computer science undergraduate
programmes~\cite{Bowen2016}.

\section{FROM Z TO THE B-METHOD}

Despite its success, Abrial became increasingly dissatisfied
with Z's limited support for systematic development and proof in
refining a Z specification to a program. In particular, he
sought a framework in which specifications could be refined step
by step into executable code, with correctness preserved by
construction.

This motivation led to the development of the B-Method in the
1980s and 1990s. B integrated specification, refinement, and
proof into a single methodological framework, supported by tools
capable of generating and discharging proof obligations.
Abrial's book \emph{The B-Book: Assigning Programs to Meanings}
\cite{JRA1996e} provided the definitive exposition of the
method.

Unlike many formal approaches, B was deliberately engineered for
industrial use. Its most prominent successes include the
development of safety-critical software for the Paris Métro
lines~14 and~1, where B-based development replaced extensive
testing with mathematical proof~\cite{Habrias2026}. It was also
used for the 1990 French population
census~\cite{BernardLaffitte1995} and by the French company
CLEARSY, e.g., for the CLEARSY Safety Platform (CSSP) in
safety-critical applications up to Safety Integrity Level 4
(SIL4)~\cite{Lecompte2020}, the higest level of functional
safety defined by the International Electrotechnical Commission
(IEC).

\section{EVENT-B AND THE RODIN PLATFORM}

In the 2000s, Abrial further generalized the B-Method into
Event-B \cite{JRA2010a}, a framework emphasizing system-level
modelling, events, and refinement. Event-B is particularly
suited to discrete transition systems, and it facilitates
reasoning at higher levels of abstraction~\cite{Hoang2013}.

From 2004 onward, Abrial played a central role in the European
RODIN project, which produced the Rodin open-source software
platform for Event-B \cite{JRA2005a}. Rodin combined automated
and interactive theorem proving with model management and
refinement support, significantly lowering the barrier to
industrial adoption~\cite{JRA2014a}. From 2004 to 2009,
Abrial served as Professor of Software Engineering at ETH~Zurich
\cite{AcademiaEuropaea}, where he influenced a new generation of
researchers in formal methods.

\section{RECOGNITION AND LATER ACTIVITIES}

Abrial's contributions have been widely recognized. He was
elected a member of the Academia Europaea in 2006
\cite{AcademiaEuropaea} and received an honorary doctorate from
the Universit\'e de Sherbrooke in 2008 \cite{Habrias2026}. He
has held visiting positions at leading institutions worldwide,
including in 2011 at Peking University in China \cite{PKU2011}
and Microsoft Research in the USA \cite{JRA2011c}.  He was also
a visitor in the School of Computer Science and Software
Engineering, at East China Normal University in Shanghai, as an
adjunct professor \cite{ECNU2017}.  Subsequently, Abrial won the
2016 Chinese National Science and Technology International
Cooperation Award, held on 9 January 2017 at the Great Hall of
the People in Beijing \cite{Kun2017}, and was congratulated by
President Xi.  His software development approach has been used
for Chinese city metro lines, including in Guangzhou, Shanghai,
and Wuhan.

Even in later years, Abrial continued to reflect critically
on the state of software engineering, advocating for greater
rigour, better integration of proof into development processes,
and stronger links between theory and industrial practice.

\section{IMPACT AND LEGACY}

Jean-Raymond Abrial's legacy lies not only in specific languages
or tools, but in a coherent vision of software engineering as a
mathematical engineering discipline. His insistence on
abstraction, refinement, and proof has influenced both academic
research and industrial practice.

The Z notation, B-Method, and Event-B remain central reference
points in formal methods, and the success of B in large-scale
safety-critical systems stands as a rare example of sustained
industrial uptake of formal techniques. Abrial's career
demonstrates that mathematical rigour and practical engineering
need not be opposed, but can be mutually reinforcing.

A wide range of tributes to Jean-Raymond Abrial from
colleagues have been published in \emph{FACS FACTS}, including
on his lesser-known exploration exploits in North Africa and
elsewhere, as well as covering his scientific achievements
\cite{HabriasBowen2025,BowenHabrias2026}, and the \emph{Formal
Aspects of Computing} journal \cite{Jones2026,Woodcock2026}.  A
blog dedicated to Jean-Raymond Abrial has been established
\cite{Blog}.  A comprehensive chronological bibliography of
J.-R.\ Abrial is also included here as an appendix and cited
where appropriate.

\section{ACKNOWLEDGMENTS}

The first author acknowledges the extensive historical material
collected by Henri Habrias, which provided an invaluable
documentary basis for and help with this biographical account
\cite{Habrias2026}. The bibliography in the appendix was
originally compiled by Henri Habrias, aided by Dominique Cansell
\cite{BowenHabrias2026,JRA2025}. The version included here has
been updated and expanded. Thank you to Egon B\"orger
\cite{JRA1996b}, Yamine Ait Ameur \cite{JRA2014a}, Troy Astarte,
Dag Spicer, David Hemmendinger, and Henry Lowood for helpful
suggestions, information, and encouragement to improve this
article. The photograph of Jean-Raymond Abrial was provided by
Dines Bj\o{}rner under a Creative Commons
license~\cite{Wikipedia}.

\def\reftitle#1{\emph{#1}}

\renewcommand{\refname}{REFERENCES}

\def\say#1{``#1''}

\section{APPENDIX}

We provide a comprehensive bibliography of
Jean-Raymond Abrial, in chronological order, by
year.

Jean-Raymond Abrial produced his first publication in 1966, when
he was working on the LTR2 (Real-Time Language) language at
the Centre de Programmation de la Marine in Paris
\cite{JRA1966}. In Grenoble, where he designed the
multidatabase system (MDBS) Socrate \cite{JRA1970a}, he
published notes on the nascent Z language and in 1974,
\emph{Data Semantics} \cite{JRA1974}, the premise of his
future work. Invited to Oxford University in 1980 by Tony
Hoare, he participated in the version of Z that became widely
distributed \cite{JRA1980a,JRA1980b,JRA1980c}.

Within the 1984 paper \emph{Specifying or how to materialise
the abstract} \cite{JRA1984} (in French), there is a
pedagogical example of his approach that led him to his B
approach and to the associated Atelier~B tool. \emph{The
B-Book} was released in 1996 \cite{JRA1996e}. With
\emph{Extending B without changing it (for developing
distributed systems)} \cite{JRA1996a}, published the same
year, and with B$^{\#}$ \cite{JRA2003c} in 2003, the
beginnings of Event-B can be seen. The book \emph{Modeling in
Event-B, System and Software Engineering} was released in 2010
\cite{JRA2010a}. The method is defined and used in many both
old and new examples that are fully developed and proven with
the Rodin tool. Then, as before, Jean-Raymond Abrial
successfully tested his method in different fields: sequential
and distributed algorithms, systems with their environment,
hybrid systems, and many mathematical theorems. Finally, he
defined an instantiation of Rodin contexts to facilitate the
modelling of theorems and their proofs.

\begin{quote}
\textit{\say{The very first paper on Z was published in 1980 (at
the time, the name Z was not `invented'), then the book on the B
method was published in 1996, and, finally, the book on Event-B
was published in 2010. So, 30 years separate Z from Event-B. It
is thus clear that I spent a significant time of my scientific
professional life working with the same kind of subject in mind,
roughly speaking specification languages. I do not know whether
this kind of addiction is good or bad, but what I know is that I
enjoyed it a lot.}} --- J.-R.\ Abrial \cite{JRA2013a}
\end{quote}

\noindent
The online DBLP database also maintains a list of Jean-Raymond
Abrial's more formal computer science publications~\cite{DBLP}.

\def\bibsection#1{\vspace{-2.5ex}\subsection*{\normalsize\bf #1}\vspace{2ex}}
\def\year#1{\vspace{-3ex}\subsection*{\normalsize\bf\em #1}\vspace{0.5ex}}

\renewcommand{\refname}{CHRONOLOGICAL BIBLIOGRAPHY}

\end{document}